\documentclass{aa}
\usepackage[T1]{fontenc}
\usepackage[colorlinks=true, linkcolor=blue]{hyperref}
\usepackage[figure, figure*]{hypcap}
\usepackage{graphicx}
\usepackage{amssymb}
\usepackage{float}
\usepackage{color}
\usepackage{geometry}
\usepackage{marginnote}
\usepackage{amsmath}
\usepackage{natbib}
\usepackage{bibentry}
\usepackage{array} 
\usepackage[varg]{txfonts}
\usepackage{caption}
\usepackage{bm}
\usepackage{subcaption}
\usepackage{xcolor}
\usepackage[super]{nth}
\usepackage[labelfont=color]{caption}
\DeclareMathSymbol{\varOmega}{\mathord}{letters}{"0A}
\DeclareMathSymbol{\varPsi}{\mathord}{letters}{"09}
\DeclareMathSymbol{\varPhi}{\mathord}{letters}{"08}
\DeclareMathSymbol{\varGamma}{\mathord}{letters}{"00}
\DeclareMathSymbol{\varPi}{\mathord}{letters}{"05}
\DeclareMathSymbol{\varLambda}{\mathord}{letters}{"03}
\DeclareCaptionFont{red}{\color{red}}
\usepackage{mathrsfs}
\usepackage[labelfont=bf]{caption}
\graphicspath{{./Plots/}}

\begin{document}

\title{Streaming instability of multiple particle species II -- \\ Numerical convergence with increasing particle number}
\author{Noemi Schaffer \inst{1}
\and Anders Johansen \inst{1,2}
\and Michiel Lambrechts \inst{1}}
\institute{Lund Observatory, Department of Astronomy and Theoretical Physics, Lund University, Box 43, 22100 Lund, Sweden \and Centre for Star and Planet Formation, Globe Institute, University  of Copenhagen, \O ster Voldgade 5–7, 1350 Copenhagen, Denmark \\ \email{noemi.schaffer@astro.lu.se}}

\date{}

\abstract{The streaming instability provides an efficient way of overcoming the growth barriers in the initial stages of the planet formation process. Considering the realistic case of a particle size distribution, the dynamics of the system is altered compared to the outcome of single size models. In order to understand the outcome of the multi-species streaming instability in detail, we perform a large parameter study in terms of particle number, particle size distribution, particle size range, initial metallicity and initial particle scale height. We study vertically stratified systems and determine the metallicity threshold for filament formation. We compare these with a system where the initial particle distribution is unstratified and find that its evolution follows that of its stratified counterpart. We find that change in particle number does not result in significant variation in the efficiency and timing of filament formation. We also see that there is no clear trend for how varying the size distribution in combination with particle size range changes the outcome of the multi-species streaming instability. Finally, we find that an initial metallicity of $Z_{\rm{init}}=0.005$ and $Z_{\rm{init}}=0.01$ both result in similar critical metallicity values for the start of filament formation. Our results show that the inclusion of a particle size distribution into streaming instability simulations, while changing the dynamics as compared to mono-disperse systems, does not result in overall unfavorable conditions for solid growth. We attribute the sub-dominant role of multiple species to the high-density conditions in the midplane, conditions under which also linear stability analysis predict little difference between single and multiple species.}


\titlerunning{Streaming instability of multiple particle species II}

\maketitle
\section{Introduction}

The initial stages of the planet formation process are hindered by various barriers. One important hurdle is the radial drift barrier, where solids of mm -- cm sizes drift as rapidly towards the star as they grow, limiting the overall particle size that can be achieved before reaching the star \citep{Birnstiel2014}. A key process in aiding solid concentration is the streaming instability \citep{Youdin2005,Youdin2007, Johansen2007, Capelo2019, Schneider2019}. When the back-reaction of the protoplanetary disk solids onto the gas is taken into account, local and nearly axisymmetric particle filaments can form. Once these reach the Roche density, their gravitational collapse produces planetesimals of 50 to a few 100 kilometers in size \citep{Johansen2015,Simon2016, Schafer2017, Klahr2020}. 

Protoplanetary disk observations and laboratory experiments that aim to recreate conditions in such settings confirm that solids with a non-monodisperse size distribution coexist in these systems. ALMA observations of the rings in the protoplanetary disk around HL Tau suggest the presence of up to centimeter sized solids \citep{ALMA, Zhang2015}. In the case of the TW Hya system, which is one of the closest observed systems with a protoplanetary disk, \cite{Menu2014} study a large set of observational data covering a wide range of near-infrared to millimeter/centimeter wavelengths, together with spectral energy distribution data. They show that models that reproduce the observations include grains that range from microns to centimeters in size. Laboratory experiments that study interactions between  porous aggregates also show that the outcome of collisions is grains of varying sizes \citep{Guttler2010, Zsom2010}.

In line with the above, many streaming instability models in recent years have included multiple particle species. \cite{Krapp2019} and \cite{Paardekooper2020} find that the linear phase of the instability has longer growth timescales than seen in mono-disperse models. However, \cite{McNally2021}, using the methodology described in \cite{Paardekooper2021}, expand the parameter space from their nominal set described in \cite{Paardekooper2020} to a larger range of particle sizes and size distributions, which are more in line with those predicted by dust coagulation/fragmentation models \citep{Birnstiel2011, Birnstiel2015}. They see that a fast growth regime is achieved given that the particle size distribution has sufficient fraction of large particles, peaks at a friction time of $0.1 \varOmega^{-1}$ and has a local dust-to-gas ratio above unity. Similarly, extending the parameter space, \cite{Zhu2020} also see that given an increased maximum solid size or total dust-to-gas ratio, the multi-species streaming instability is indeed favorable for solid growth. The non-linear phases of the instability were studied in \cite{Bai2010} and \cite{Schaffer2018}. Given super-solar metallicity, \cite{Bai2010} see the formation of dense filament clumps and in \cite{Bai2010a} show that one particle per grid cell is sufficient for numerical convergence.

In this paper, we build on the results presented in \cite{Schaffer2018}. We study how increasing the particle number, considering various particle size distributions and varying the particle size range, affects the efficiency of the streaming instability. Our goal is to find the critical metallicities where particle filaments begin to form in order to compare how the previously mentioned parameters influence particle clumping through the streaming instability. We aim to understand if the particle number plays as strong a role in the efficiency of the multi-species instability in the stratified case, as reported by \cite{Krapp2019} for the unstratified case.

\section{Numerical method}

Using the Pencil Code\footnote{Publicly available at http://pencil-code.nordita.org/} we perform two-dimensional shearing box simulations that represent a local segment of a protoplanetary disk. Our system co-rotates the central star at a distance $r$ with a Keplerian velocity of $\textbf{\textit{v}}_{\rm{K}}$. Our coordinate axes are oriented with
$x$ pointing radially outward, $y$ along direction of the disk rotation and $z$ vertically out of the disk. We neglect both magnetic effects and the self-gravity of the solids. The equations that drive the gas component are the momentum and continuity equations, which are presented in Eqs. 1 and 2 of \cite{Schaffer2018}.



Here, the gas density and velocity are defined as $\rho_{\rm{g}}$ and $\textbf{\textit{{u}}}$, respectively. The local Keplerian angular frequency is $\varOmega$ and the dust-to-gas ratio is defined as $\varepsilon=\rho_{\rm{p}}/\rho_{\rm{g}}$, with $\rho_{\rm{p}}$ denoting the particle density. The dimensionless parameter that sets the large-scale pressure gradient of the disk is defined as

\begin{equation}
\eta = - \frac{1}{2} \Bigg(\frac{H_{\rm{g}}}{r}\Bigg)^2 \frac{\partial \mbox{ln } P}{\partial \mbox{ln } r}.
\label{eta}
\end{equation}

\noindent As in \cite{Schaffer2018}, we non-dimensionalize the pressure gradient as $\varPi = \eta v_{\rm{K}}/c_{\rm{s}}$, where $c_{\rm{s}}$ is the sounds speed, and choose $\varPi =0.05$, as it represents well the pressure support in the inner protoplanetary disk \citep{Bai2010}. In Eq. \ref{eta}, $H_{\rm{g}}$ is the gas scale height and $P$ the pressure. Finally, $\tau_{\rm{f}}$ is the friction time of the solid component, which in the Epstein regime is defined as 

\begin{equation}
\tau_{\rm{f}} = \frac{\rho_{\bullet} a}{\rho_{\rm{g}} c_{\rm{s}}},
\label{tauf}
\end{equation}

\noindent where $\rho_{\bullet}$ is the internal solid density and $a$ the solid radius. Using $\varOmega$ to non-dimensionalize $\tau_{\rm{f}}$ gives the Stokes number, which becomes 

\begin{equation}
\tau_{\rm{s}} = \sqrt{2 \pi} \frac{\rho_{\bullet} a}{\varSigma_{\rm{g}}},
\label{taus}
\end{equation}

\noindent where $\varSigma_{\rm{g}}$ is the gas surface density. In the rest of paper, we use the Stokes number to represent particle sizes.

The solid component is modeled using super-particles, which represent swarms of physical particles. These follow the equations presented in Eqs. 4 and 5 of \cite{Schaffer2018}. The individual particles have velocity $\textit{\textbf{v}}_i$ and position $\textit{\textbf{x}}_i$. The velocity dispersion of the superparticles is captured as the distribution of velocities between the superparticles present within each grid cell. The back-reaction friction force from the particles onto the gas is calculated by mapping the particle density $\rho_{\rm{p}}$ onto the grid using a TSC (Triangular Shaped Cloud) scheme \citep{Youdin2007}.




The initial conditions for the gas and particle density profile both follow a stratified profile centered around the midplane. The nominal initial particle scale height is $H_{\rm{p, init}} = 0.015 H_{\rm{g}}$. In order to ensure that the initial particle scale height does not influence the results, we also perform a run with $H_{\rm{p, init}} = 1 H_{\rm{g}}$. In this case the initial particle profile becomes unstratified within our simulation domain. The gas component initially has a sub-Keplerian velocity, such that $\textbf{\textit{u}}(t = 0) =  - \eta \textit{v}_{\rm{K}}\hat{\textit{\textbf{y}}}$ and the initial particle velocity is zero. 

We fix the simulation domain at $0.2 H_{\rm{g}} \times 0.2 H_{\rm{g}}$ with a resolution of 128 $\times$ 128 grid cells. The boundary conditions in the vertical direction are reflecting. Some small particles hit the vertical boundary and become reflected, however following our findings in \cite{Schaffer2018}, we do not expect this to have a significant effect on the results. The boundary conditions in the radial direction are periodic.

To understand how the streaming instability evolves when multiple particle species are considered, instead of performing several runs with different starting metallicities, we remove gas from the system \citep{Carrera2015}. The nominal initial metallicity we implement is $Z_{\rm{init}}=0.01$. In addition, we also perform a run with $Z_{\rm{init}}=0.005$ to ensure that our choice for this value does not influence the outcome. We start removing gas after the first 30 orbits, at which time the particles have saturated to a constant scale height. We decrease the gas density in an exponential manner, such that it is halved every 50 orbits, following

\begin{figure}[t!]
\centering
\includegraphics[width=1\columnwidth]{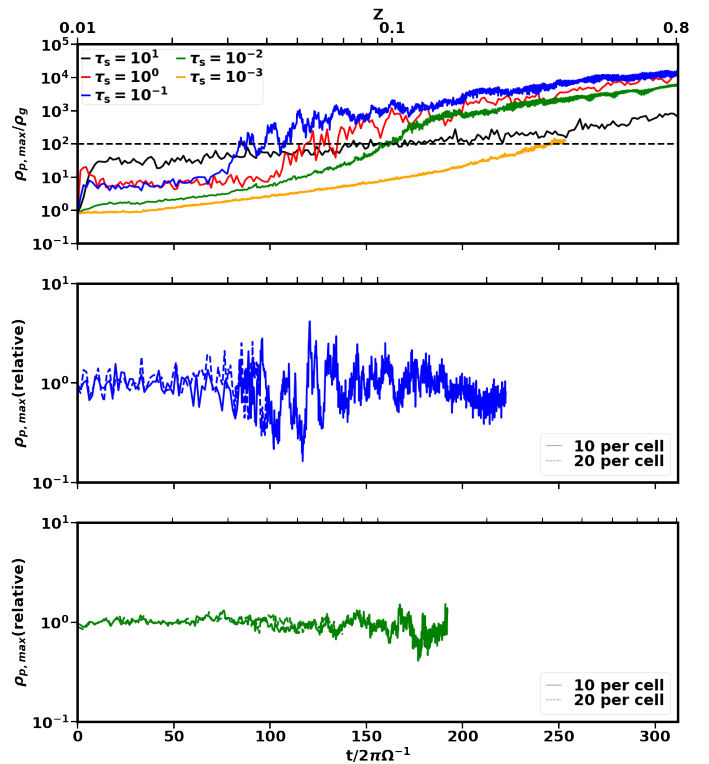}
\caption{The top panel shows the maximum solid density evolution as a function of time for single species runs with $\tau_{\rm{s}} = 10^{-3}, 10^{-2}, 10^{-1}, 10^{0} \text{ and } 10^{1}$. The metallicity is shown on the top axis. The horizontal black dashed line corresponds to $\rho_{\rm{p,max}}/\rho_{\rm{g}}=100$, which represents our adopted metallicity threshold for filament formation. The middle and bottom panels show the effect of changes in $N_{\rm{p}}$ for $\tau_{\rm{s}} = 10^{-1} \mbox{ and } 10^{-2}$, respectively. Here, the solid lines show the relative maximum density for 10 vs 20 particles per cell while the dashed lines show the relative maximum density for 10 vs 100 particles per cell. In both cases, the level of deviation from unity is low, implying that the particle number does not have a significant effect on the maximum density of the single-species streaming instability. The colors correspond to the legend in the top panel. }
\label{rhopmax_single}
\end{figure}

\begin{equation}
\rho_{\rm{g}}(t) = \rho_{\rm{g,0}} \exp(-kt).
\end{equation}

\noindent Here, $\rho_{\rm{g,0}}$ is the initial gas density and $k=1/\tau$, where $\tau$ is the e-folding time and the half-life time is defined as $t_{\rm{rem}} = \tau / \ln(2)$. We also perform convergence tests, such that the gas density is halved every $t_{\rm{rem}} =  100 \mbox{ }$ orbits instead (see Appendix \ref{App1}). As the particle density is kept constant overall, this results in continuous metallicity increase with time, where the metallicity is defined as

\begin{equation}
Z = \bar{\varSigma_{\rm{p}}}/\bar{\varSigma_{\rm{g}}},
\end{equation}

\noindent with $\bar{\varSigma_{\rm{p}}}$ as the particle surface density.

\section{Results}

\subsection{Single-species validation}
\label{SingleSection}

\begin{figure}[t!]
\centering
\includegraphics[width=1\columnwidth]{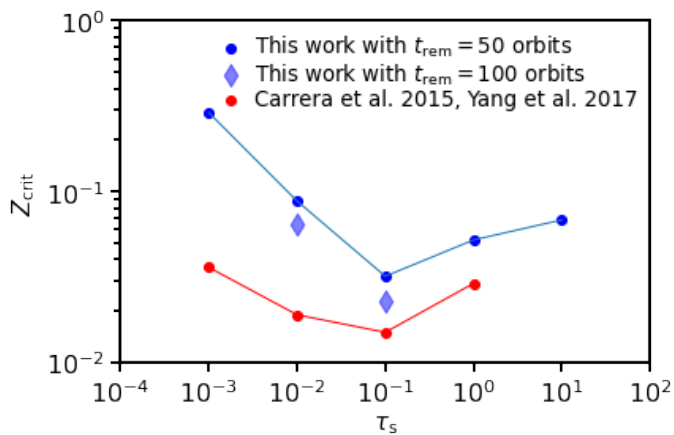}
\caption{Critical metallicity threshold for filament formation, $Z_{\rm{crit}}$, with respect to Stokes number for gas removal times of 50 and 100 orbits. With red we show the critical metallicity values measured by \cite{Carrera2015} and \cite{Yang2017} for comparison. These values are somewhat lower due to differences in initial conditions, resolution and in the definitions of the start of filament formation.}
\label{Zclump_single_compare}
\end{figure}

In order to determine the right criteria for the start of filament formation, we begin with single-species simulations. We then compare the results with that of \cite{Carrera2015} and \cite{Yang2017} to ensure that our metallicity limit, $Z_{\rm{crit}}$, holds up against the expectations. 

Our criteria for the start of filament formation is $\rho_{\rm{p,max}}/{\rho}_{\rm{g}} = 100$, where $\rho_{\rm{p,max}}$ is the maximum solid density. This criteria is an order of magnitude approximation for the Roche density at 2.5 au in a MMSN disk \citep{Yang2017}. On the top panel of Fig. \ref{rhopmax_single}, we show the evolution of the maximum solid density with respect to the gas density of all single species runs, where $\tau_{\rm{s}} = 10^{-3},10^{-2},10^{-1},10^{0},  \text{ and } 10^{1}$. The dashed horizontal line corresponds to $\rho_{\rm{p,max}}/{\rho}_{\rm{g}} = 100$. As seen on the first panel of Fig. \ref{rhopmax_single}, all five particle species reach our prescribed limit. The curve corresponding to $\tau_{\rm{s}}=10^{-3}$ has the slowest maximum solid density increase and little small-scale variations, which suggests that the solids sediment to a dense midplane layer without the emergence of significant clumping.

We also study the importance of the numerical sampling effects by performing runs with the same parameters while increasing the particle number from $N_{\rm{p}} = 10$ to $N_{\rm{p}} = 20$ and to $N_{\rm{p}} = 100$ per cell. The result of this is shown for $\tau_{\rm{s,max}} = 10^{-1}$ on the middle and for $\tau_{\rm{s,max}} = 10^{-2}$ on the bottom panel of Fig. \ref{rhopmax_single}. Here, we see that both the curves corresponding to $N_{\rm{p}} = 10$ vs $N_{\rm{p}} = 20$ and $N_{\rm{p}} = 10$ vs $N_{\rm{p}} = 100$ per cell remain close to unity. The largest variations from unity coincide with the times when the maximum solid density curves shown on the top panel cross the $\rho_{\rm{p,max}}/{\rho}_{\rm{g}} = 100$ limit. Afterwards, the level of deviation settles down to close to unity again. This implies that numerical parameters such as particle number do not have a significant effect on the evolution of the single-species streaming instability.

In Appendix \ref{Appendix1}, we present the space-time diagrams of the single species runs, where $\tau_{\rm{s}} = 10^{-3},10^{-2},10^{-1},10^{0},  \text{ and } 10^{1}$ and $N_{\rm{p}}=10 $ per cell. As show in Fig. \ref{single_space_time} by the orange horizontal lines, the threshold metallicity for clumping, $Z_{\rm{crit}}$, is reached in all five cases. The figure also shows that indeed, the filaments above the $Z_{\rm{crit}}$ limit are clearly visible and permanent, except in the case of the first panel, where $\tau_{\rm{s}} = 10^{-3}$. Here, the generated turbulence is not strong enough and the particles settle into a thin, dense layer thus increasing the solid density without the formation of large filaments. As Fig. \ref{uz_0p001} shows, here we see evidence of some vertically-elongated structures in the vertical component of the gas velocity, which could be the result of the vertical shear instability \citep{Lin2021}. However, overall we see that our $Z_{\rm{crit}}$ limit catches the particle filaments well, so we apply this criteria moving forward. Also, we prescribe this criteria not necessarily to give absolute values for solid clumping, but rather to allow us to make comparisons between the importance that various parameters play in the streaming instability. As high resolution runs are not computationally feasible for the large parameter study presented in this paper, we focus on determining the relative differences in $Z_{\rm{crit}}$ and thus the parameter space in question.

In Fig. \ref{Zclump_single_compare} we summarize the $Z_{\rm{crit}}$ values for all five particle species. We also show the critical metallicity values presented in \cite{Carrera2015} and \cite{Yang2017} with red and see that our values are somewhat higher. This is especially true in the case of $\tau_{\rm{s}} = 10^{-3}$ and $10^{-2}$. One reason for the discrepancy is likely that we set the initial metallicity to $Z_{\rm{init}} = 0.01$, whereas in both \cite{Carrera2015} and \cite{Yang2017}  $Z_{\rm{init}} = 0.005$. In addition, we also have a different way of defining the beginning of filament formation. \cite{Carrera2015} define particle clump\textbf{s} as temporally stable overdensities in the solid surface density, while \cite{Yang2017} perform high resolution studies with a fixed metallicity to find the critical metallicity by tracking the maximum local solid density. In Fig. \ref{Zclump_single_compare} we also show the effect of doubling the gas removal time, $t_{\rm{rem}}$. In the case of both $\tau_{\rm{s}} = 10^{-1}$ and $\tau_{\rm{s}} = 10^{-2}$ the change in $Z_{\rm{crit}}$ is about a factor of one half and in the direction towards the results of \cite{Carrera2015} and \cite{Yang2017}.

\subsection{Multi-species results}

\begin{figure}[!t]
\centering
\includegraphics[width=1\columnwidth]{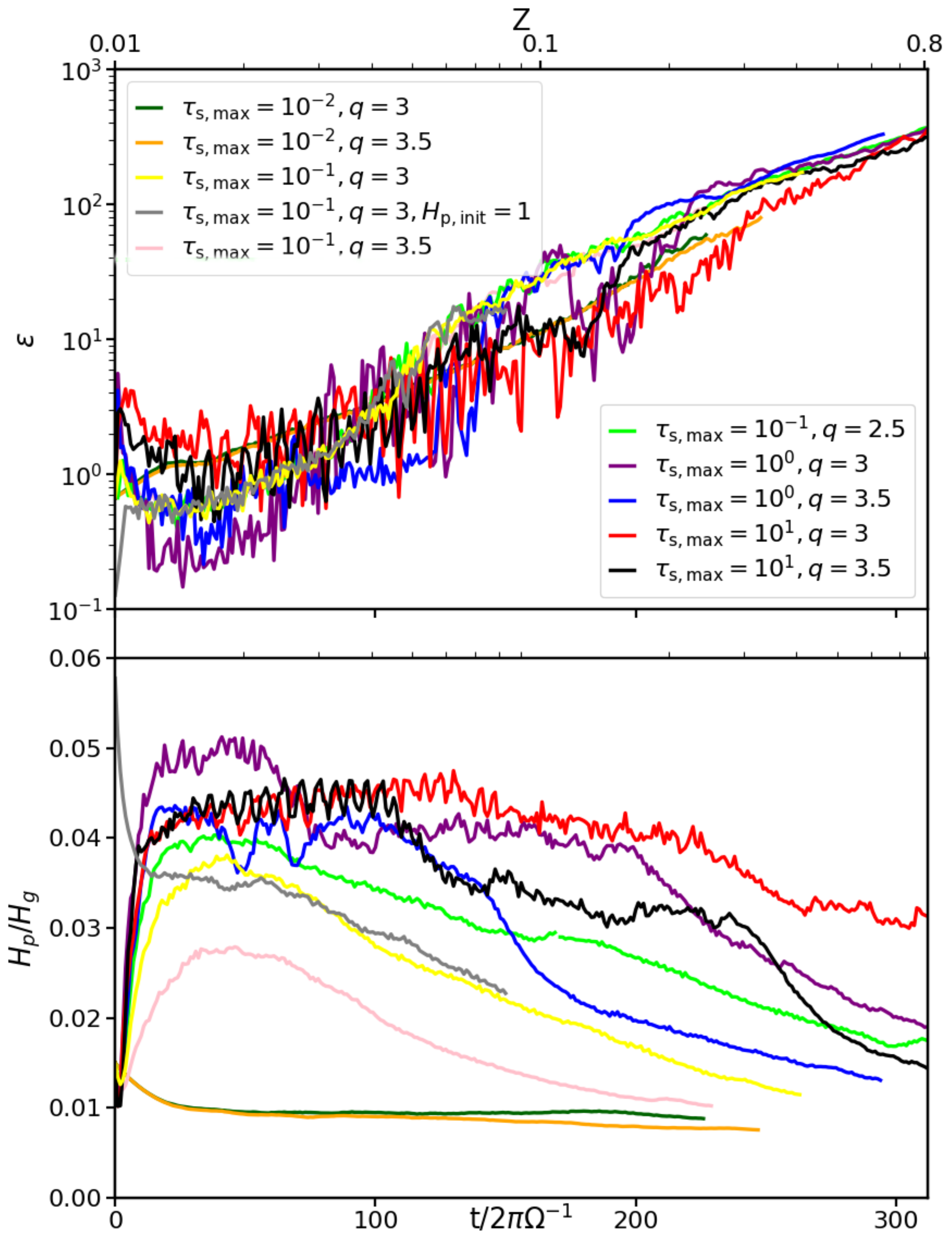}
\caption{Average dust-to-gas ratio in the midplane (top panel) and particle scale height as a function of gas scale height (bottom panel) with respect to time for all Stokes number ranges and size distribution exponents with $N_{\rm{p}} = 70$ per cell. The initial particle scale height is $H_{\rm{p, init}} = 0.015 H_{\rm{g}}$, except in the case of the grey curves, where $H_{\rm{p, init}} = 1 H_{\rm{g}}$. The metallicity is shown on the top axis. The top panel shows that $\epsilon$ is not significantly affected by the slope of the size distribution, but the short-time variations increase with $\tau_{\rm{s,max}}$. The bottom panel shows that the particle scale heights saturate closer to the midplane for small size ranges, but there is no clear trend for how $H_{\rm{p}}/H_{\rm{g}}$ changes as a function of $q$ and $\tau_{\rm{s,max}}$.}
\label{compare_epsilon}
\end{figure}

In this section, we study how the streaming instability behaves once multiple particle species are considered. We investigate the significance of the number of simulation particles, $N_{\rm{p}}$, maximum Stokes number, $\tau_{\rm{s, max}}$, solid size distribution exponent, $q$, initial metallicity, $Z_{\rm{init}}$, and initial particle scale height, $H_{\rm{p, init}}$, on the resulting metallicity threshold, $Z_{\rm{crit}}$. 

We summarize the run parameters in Table \ref{MultiTable}. The first two columns show the minimum and maximum values of the particle size range. Between these values, the particles are distributed in a continuous way, such that each particle in a given run has a unique size chosen randomly to sample the assumed size distribution. The third column shows $q$, the exponent of the particle size distribution described as $\rm{d}\it{N} / \rm{d}\it{a} \propto \it{a^{-q}}$. Here, $N$ is the particle number density and $a$ the particle size. The value of $q$ can take up different values, based on the physical processes considered. In this paper, we consider $q=3.5$, which applies in the case of self-similar fragmentation cascade \citep{Dohnanyi1969, Williams1994}. In addition, in order to stay consistent with various possible outcomes of solid-gas dynamics, we also implement solid size distribution exponents of $q=2.5$ and $q=3$ \citep{Birnstiel2011}. The fourth column in Table \ref{MultiTable} shows the number of simulation particles, $N_{\rm{p}}$, per cell. The fifth column shows the initial metallicity, $Z_{\rm{init}}$  and the final column the initial particle scale height, $H_{\rm{p, init}}$, of the system.

\begin{figure}[!t]
\centering
\includegraphics[width=1\columnwidth]{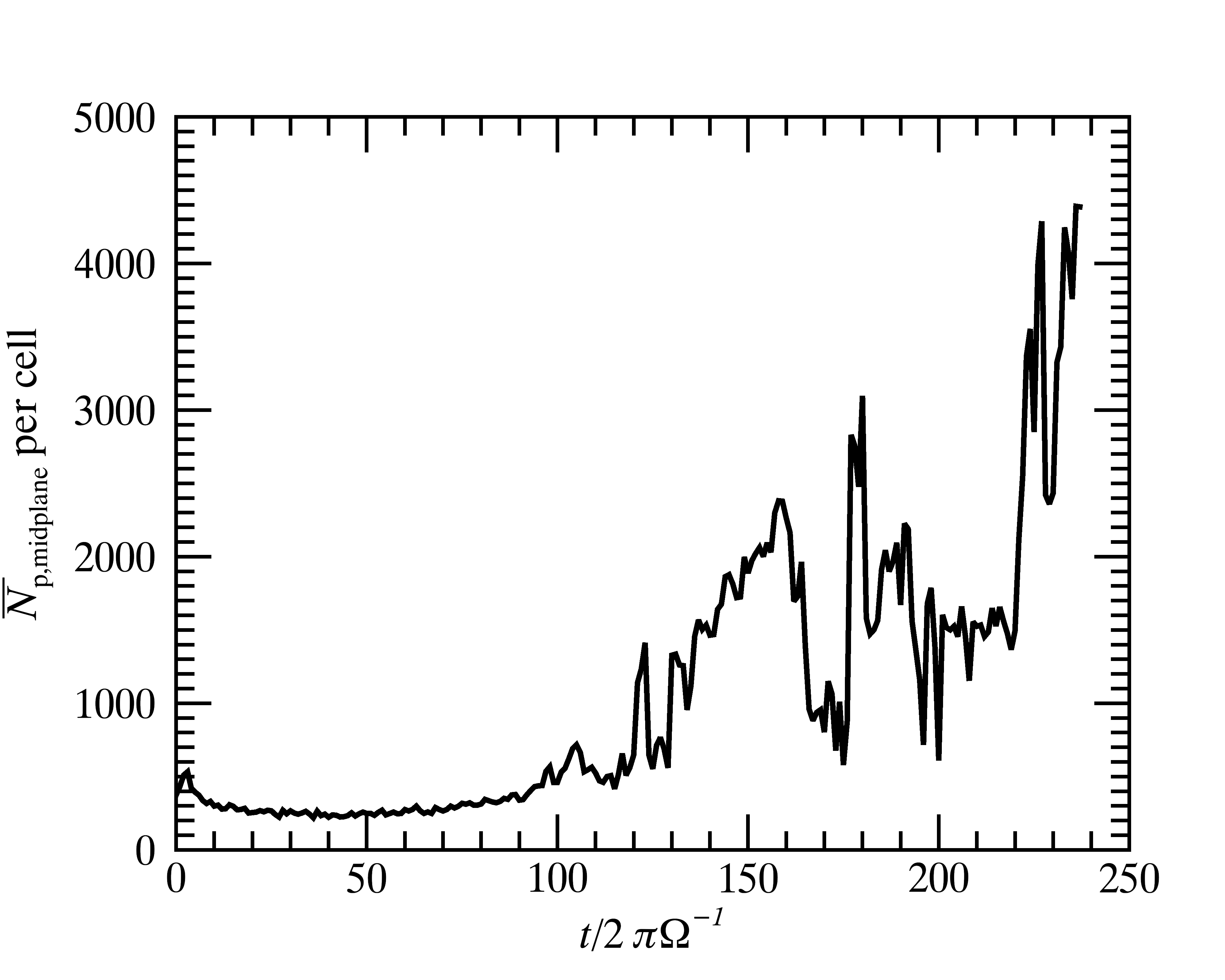}
\caption{Average particle number per cell in the midplane as a function of time. Initially, the particle number is $N_{\rm{p}}=70$ per cell, the particle size range is $\tau_{\rm{s}} = 10^{-3} - 10^{-1}$ and the size distribution exponent is $q=3.5$. As the particles sediment towards the midplane, the particle number and thus the dust-to-gas ratio increases here significantly. The particle number is around 300 per cell, even in the early stages of the simulation. This is much higher than the mean particle number per cell, since the scale height of the dust layer is much smaller than the vertical extent of the box.}
\label{part_midpl}
\end{figure}

\begin{table}
\centering
\caption{Simulation parameters for multi-species runs}
\label{MultiTable}
\begin{tabular}{>{$}c<{$}>{$}c<{$}>{$}c<{$}>{$}c<{$}>{$}c<{$}>{$}c<{$}>{$}c<{$}}
\hline
\hline
 \tau_{\rm{s, min}} & \tau_{\rm{s, max}} & q & N_{\rm{p}}/ \text{cell} & \textit{Z}_{\rm{init}} & \textit{H}_{\rm{p, init}}/H_{\rm{g}} \\
  \hline
 10^{-3} & 10^{-1} & 2.5  & 70 & 0.01 & 0.015\\
 10^{-3} & 10^{-1} & 3 & 7 & 0.01 & 0.015\\
 10^{-3} & 10^{-1} & 3 & 7 & 0.005 & 0.015\\
 10^{-3} & 10^{-1} & 3 & 7 & 0.01 & 1\\
 10^{-3} & 10^{-1} & 3  & 70 & 0.01 & 0.015\\
 10^{-3} & 10^{-1} & 3.5 & 7 & 0.01 & 0.015\\
 10^{-3} & 10^{-1} & 3.5  & 70 & 0.01 & 0.015\\
 10^{-3} & 10^{-1} & 3.5  & 140 & 0.01 & 0.015\\
 10^{-3} & 10^{-1} & 3.5  & 280 & 0.01 & 0.015\\
 10^{-3} & 10^{-1} & 3.5  & 700 & 0.01 & 0.015\\
 \dagger 10^{-3} & 10^{-1} & 3.5  & 700 & 0.01 & 0.015\\
 \hline
 10^{-3} & 10^{-2} & 3  & 70 & 0.01 & 0.015\\
 10^{-3} &  10^{-2} & 3.5 & 70 & 0.01 & 0.015\\
 \hline
 10^{-3} &  10^{0} & 3  & 70 & 0.01 & 0.015 \\
 10^{-3} &  10^{0} & 3.5  & 70 & 0.01 & 0.015\\
 \hline
 10^{-3} & 10^{1} & 3  & 70 & 0.01 & 0.015\\
 10^{-3} & 10^{1} & 3.5  & 70 & 0.01 & 0.015\\
\hline
\hline
\end{tabular}
\end{table}

\subsubsection{Particle settling}

\begin{figure*}[!th]
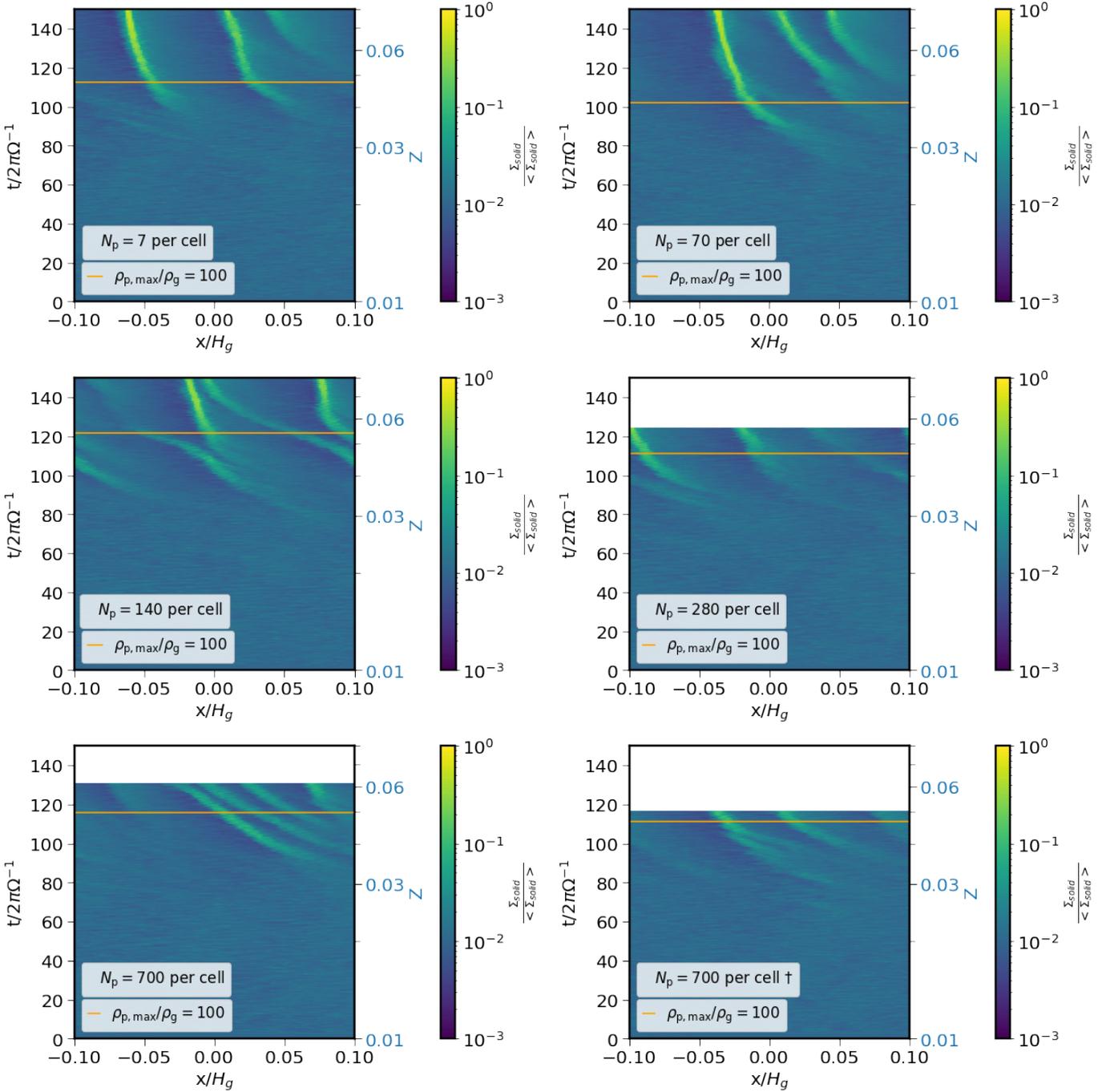

\centering
\includegraphics[width=0.49\textwidth]{filaments_q3p5_7.pdf}
\includegraphics[width=0.49\textwidth]{filaments_q3p5_contour.pdf}
\includegraphics[width=0.49\textwidth]{filaments_q3p5_x20.pdf}
\includegraphics[width=0.49\textwidth]{filaments_280.pdf}
\includegraphics[width=0.49\textwidth]{filaments_x100_contour2.pdf}
\includegraphics[width=0.49\textwidth]{filaments_700_rand.pdf}
\caption{Space-time plots showing the solid surface density for runs with $q=3.5$, $\tau_{\rm{s}} = 10^{-3}-10^{-1}$ and $N_{\rm{p}} = 7, 70, 140, 280$ vs $700$ per cell. The last panel, indicated with a dagger, corresponds to the run where $N_{\rm{p}} = 700$ per cell and the particles are initialized with a different randomization as in our nominal 700 per cell model, shown on the left-hand side of the last row. The orange line marks the limit of threshold metallicity, $Z_{\rm{crit}}$, where $\rho_{\rm{p,max}}/\rho_{\rm{g}} = 100$. The y axis on the right-hand side shows the metallicity, Z. In all four systems, the critical metallicity is reached after about 100 -- 120 orbits, signifying that increasing the particle number does not make the multi-species streaming instability less efficient in forming filaments.}
\label{filaments_q3p5_compare}
\end{figure*}

In order to understand how the multi-species streaming instability behaves, we begin by tracking the evolution of the midplane dust-to-gas ratio and the average particle scale height as a function of gas scale height. We present the evolution of $\epsilon = \rho_{\rm{p}}/\rho_{\rm{g}}$ and $H_{\rm{p}}/H_{\rm{g}}$ in the top and bottom panels of Fig. \ref{compare_epsilon}, respectively. The top panel shows that changes in either $q$ or $\tau_{\rm{s,max}}$ do not affect the evolution of the dust-to-gas ratio significantly, even though all these systems saturate to different scale heights. On Fig. \ref{part_midpl} we show the average particle number per cell in the midplane for the system corresponding to the pink curves in Fig. \ref{compare_epsilon}. We see that as the particles sediment towards the midplane, the particle number increases here significantly compared to the initial value of $N_{\rm{p}} = 70$ per cell. This shows that the particle number per cell in the mid-plane where the streaming instability evolves is much higher than the mean particle number per cell over the box. Other than our nominal initial particle scale height of $H_{\rm{p, init}} = 0.015 H_{\rm{g}}$, we also show here the model, where we set $H_{\rm{p, init}} = 1 H_{\rm{g}}$. The top and bottom panel of Fig. \ref{compare_epsilon} show that both the average dust-to-gas ratio in the midplane and the particle scale height evolve similarly, given both initial particle scale height values. Thus, we find that both an initially stratified and unstratified particle profile produce the same outcome.

There is however some deviation in how both $\epsilon$ and $H_{\rm{p}}/H_{\rm{g}}$ evolve as a function of the maximum Stokes number and the slope of the particle size distribution. In terms of the particle size range, the systems with $\tau_{\rm{s}}=10^{-3}-10^{-2}$ show the least amount of short-term variations in the development of their dust-to-gas ratios. The particle scale height in both the $q=3$ and $q=3.5$ systems implies that the particles slowly sink towards the midplane and are not stirred significantly.

Given our nominal particle size range of $\tau_{\rm{s}}=10^{-3}-10^{-1}$, there is no significant deviation in how the dust-to-gas ratio evolves based on whether $q=2.5, 3$ or $3.5$. However, the particle scale height corresponding to the system where $q=3.5$ shows closer settling to the midplane, while the system with $q=2.5$ tends to saturate farther away from the midplane. The smaller the value of the size distribution exponent, the more mass there is in the larger solids. This implies that the system with $q=2.5$ results in more turbulence than the system with $q=3$ and even more than the one with $q=3.5$, meaning that the particles are more efficiently diffused to larger scale heights.

The evolution of $\epsilon$ does not change significantly, whether the maximum particle size is $\tau_{\rm{s, max}}=10^{0}$ or $\tau_{\rm{s, max}}=10^{1}$ for either $q=3$ or $q=3.5$. In all four systems, there is quite a lot of short-term variations in $\epsilon$ during the first approximately 200 orbits, which then levels out as the particles settle closer to the midplane. Overall, the level of short-term variations in the dust-to-gas ratio seems to increase with the maximum Stokes number.

\subsubsection{Influence of particle number}

\begin{figure}[!t]
\centering
\includegraphics[width=1\columnwidth]{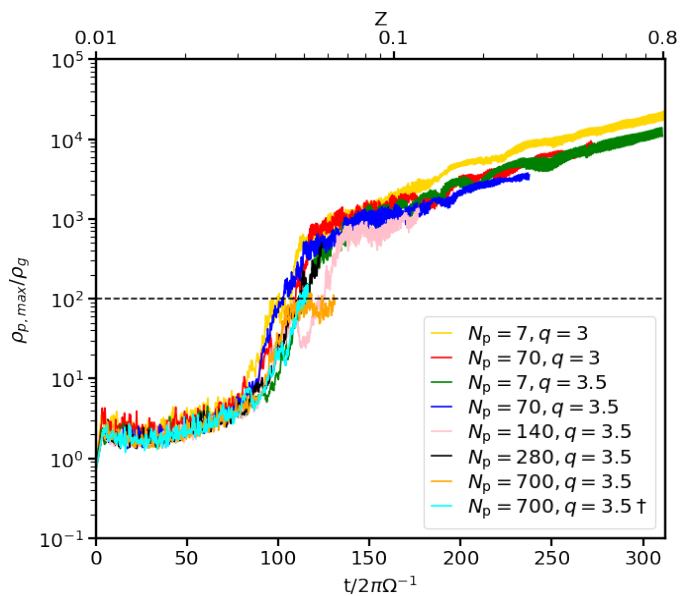}
\caption{Maximum solid density evolution as a function of time comparing the effect of particle number for runs with $\tau_{\rm{s}} = 10^{-3}-10^{-1}$ and $q=3$, $q=3.5$. The dagger symbol notes the model where we initialized the particle position with a different randomization than that of the system corresponding to the orange curve, with the same particle number.} The metallicity is shown on the top axis. The dashed line corresponds to $\rho_{\rm{p,max}}/{\rho}_{\rm{g}} = 100$, which represents the limit of solid clumping. All curves show similar trends, independent of the particle number and the slope of the solid size distribution.
\label{compare_surfdens_N}
\end{figure}

Here, we look at how the particle number affects the efficiency of filament formation.
In the space-time plots of Fig. \ref{filaments_q3p5_compare} we compare how $N_{\rm{p}} = 7, 70, 140, 280$ versus $N_{\rm{p}} = 700$ per cell changes the solid surface density evolution of systems with a size distribution exponent of $q=3.5$ and particle size range of $\tau_{\rm{s}} = 10^{-3}-10^{-1}$. We see that the critical metallicity is reached after approximately 100 -- 120 orbits in all four systems, independent of the particle number. Figure \ref{compare_surfdens_N} shows that the maximum particle density of these systems follows very similar trends as well. In both the case of the $N_{\rm{p}} = 140$ (pink curve) and the nominal $N_{\rm{p}} = 700$ per cell (orange curve) runs, there is a small dip in $\rho_{\rm{p,max}}/\rho_{\rm{g}}$ at around 100 orbits, but this is followed by increase again. We note, that the saturation of the maximum particle density seems to be somewhat delayed as the particle number is increased to $N_{\rm{p}} = 140$ per cell (pink curve) and $N_{\rm{p}} = 700$ per cell (orange curve). On Fig. \ref{compare_surfdens_N} we also show the maximum solid density evolution of the systems where $q=3$. Here, we compare $N_{\rm{p}} = 7$ with $N_{\rm{p}} = 70$ per cell and find that there is no significant difference between the corresponding curves.

Overall, Fig. \ref{compare_surfdens_N} also shows that given the fixed maximum Stokes number of $\tau_{\rm{s, max}} = 10^{-1}$, neither the particle number, nor the size distribution exponent makes a significant difference on when the limit for solid clumping is reached. We also note that the slope of the curves presented here is set not only by the efficiency of filament formation but also by the continuous removal of gas, through which we increase the metallicity.

\begin{figure*}[!th]
\centering
\includegraphics[width=1\textwidth]{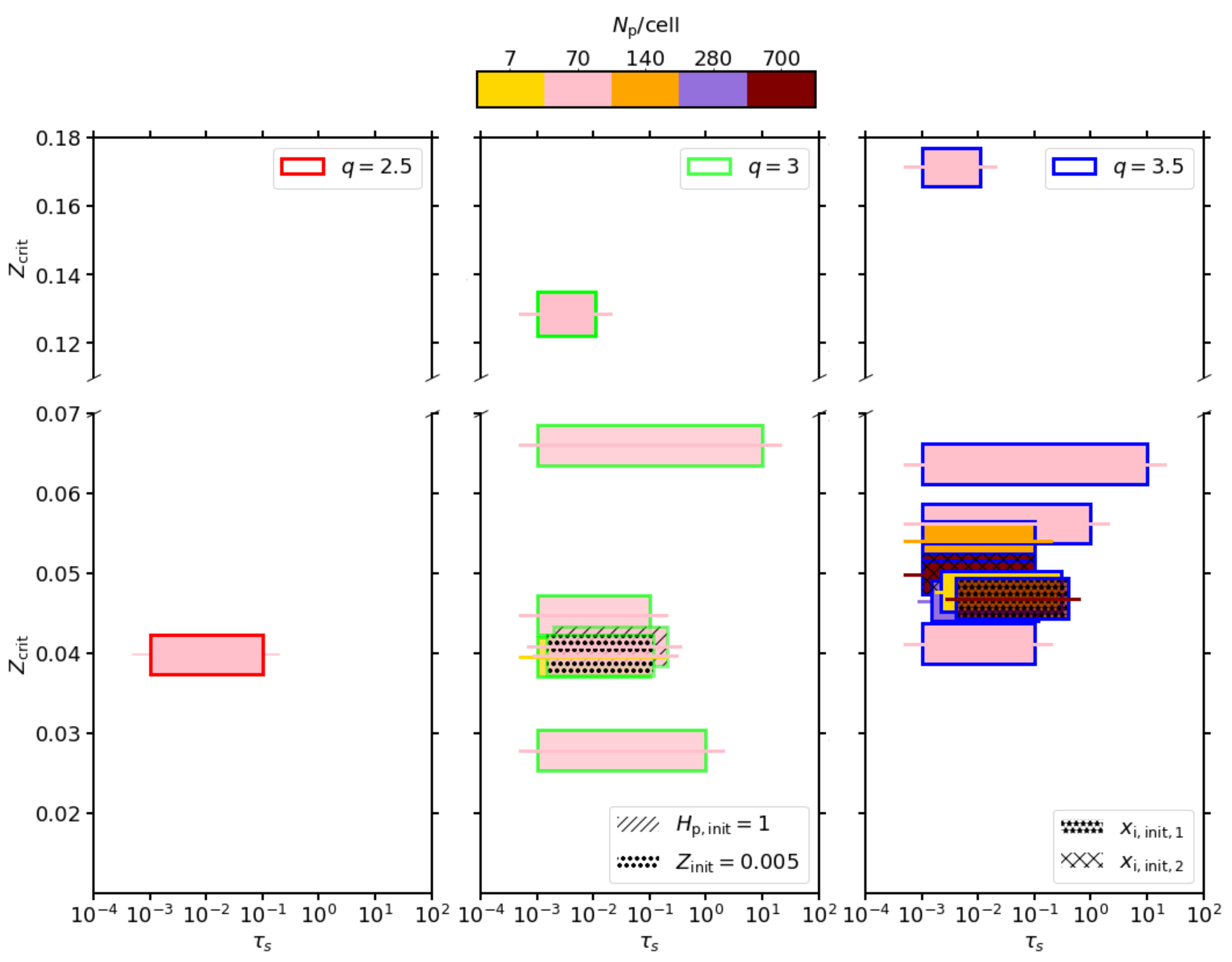}
\caption{Critical metallicity for filament formation, $Z_{\rm{crit}}$, as a function of Stokes number. The initial metallicity and particle scale height are $Z_{\rm{init}}=0.01$ and $H_{\rm{p, init}}=0.015$, unless otherwise specified. The colorbar corresponds to the number of particles used in the given simulation and the color of the box edges to the exponent of the size distribution, as shown in the legend. Some of the bar are shifted horizontally for better visibility. The legend on the left-most panel marks the runs, where the particle positions were initialized using different randomization. The horizontal lines cutting through the boxes represent the actual $Z_{\rm{crit}}$ values.}
\label{Z_clump}
\end{figure*}

To show the role of the particle number in the efficiency of filament formation, we present Fig. \ref{Z_clump}. The first panel shows the critical metallicity given $q=2.5$ for our nominal particle number of 70 per cell. The second panel corresponds to the case of $q=3$, where we find that a factor of ten increase from 7 to 70 particles per cell results in only about a 0.005 increase in $Z_{\rm{crit}}$. In the third panel we show the systems with $q=3.5$, where we see that increasing the particle number does not automatically mean an increase in $Z_{\rm{crit}}$. The threshold metallicity for filament formation is increased for 7 compared to 70 particles per cell. As we increase the particle number to 140 per cell, $Z_{\rm{crit}}$ is increased slightly but once $N_{\rm{p}}=280$ per cell, the critical metallicity is decreased again. Finally, when the particle number is increased to 700 per cell, the critical metallicity remains in the same range as in the case of fewer particles per cell, regardless of the method we choose to initialize the particle positions. Most importantly, $Z_{\rm{crit}}$ changes by only about a factor of unity as $N_{\rm{p}}$ changes by a factor of one hundred. All this implies that numerical parameters, such as particle number, do not have a significant effect on the time of filament formation, and thus the evolution of the streaming instability. We note here, that these metallicity values are calculated given our criteria that solid clumping begins once $\rho_{\rm{p,max}}/{\rho}_{\rm{g}} = 100$.

\begin{figure}[!th]
\centering
\includegraphics[width=0.7\columnwidth]{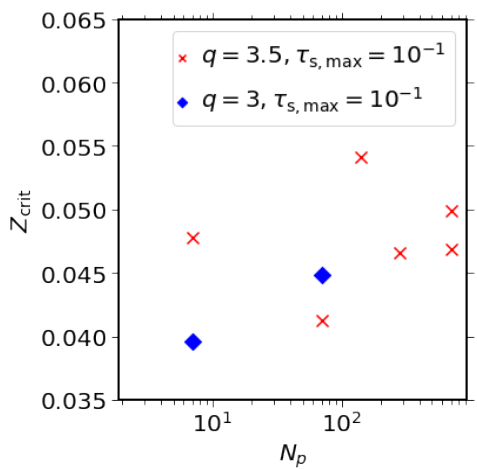}
\caption{Critical metallicity for filament formation, $Z_{\rm{crit}}$, as a function of particle number. Here, $\tau_{\rm{s, max}}$ is fixed as $10^{-1}$, while $q=3$ and $q=3.5$. There is no significant change in $Z_{\rm{crit}}$ with change in particle number for either size distribution exponent.}
\label{Z_clump_N}
\end{figure}

We summarize the importance of the particle number in Fig. \ref{Z_clump_N}. We clearly see that given $q=3 \text{ and } 3.5$, $N_{\rm{p}}$ does not have a significant effect on the critical metallicity of particle clumping and that an increase in particle number does not directly translate to higher $Z_{\rm{crit}}$.

\subsubsection{Influence of particle size range}

\begin{figure}[!th]
\centering
\includegraphics[width=1\columnwidth]{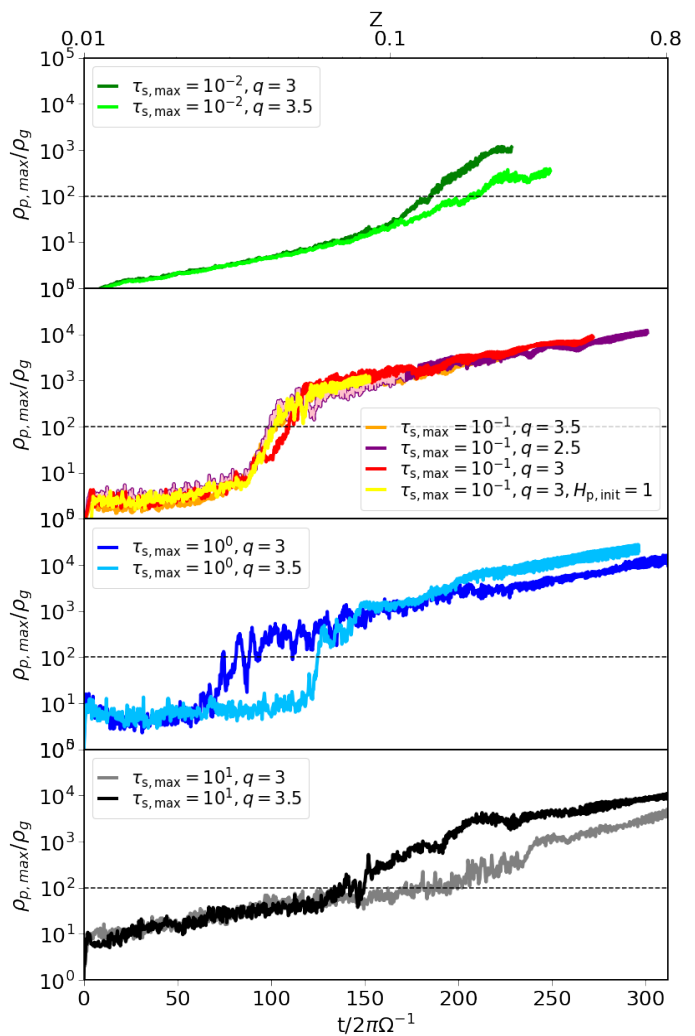}
\caption{Maximum solid density evolution as a function of time comparing the effect of the size distribution exponents and the particle size ranges given $N_{\rm{p}} = 70$ per cell. The initial particle scale height is $H_{\rm{p, init}} = 0.015 H_{\rm{g}}$, except in the case of the yellow curve, where $H_{\rm{p, init}} = 1 H_{\rm{g}}$. The metallicity is shown on the top axis. The dashed line corresponds to $\rho_{\rm{p,max}}/{\rho}_{\rm{g}} = 100$, which represents our assumed threshold for solid clumping. Independent of $q$, the maximum solid density evolves similarly in the first, second and last panels until the critical metallicity limit is reached. Given that $\tau_{\rm{s,max}} = 10^0$ however, there is a deviation of about 50 orbits between the time of filament formation, depending on whether $q=3$ or $q=3.5$. There is also no clear trend on how $\rho_{\rm{p,max}}/{\rho}_{\rm{g}}$ changes as both a function of $q$ and $\tau_{\rm{s,max}}$. }
\label{compare_surfdens_tau}
\end{figure}

We also examine the effect of the size distribution on the clumping efficiency for different particle size ranges. Figure \ref{compare_surfdens_tau} compares the evolution of runs with $\tau_{\rm{s,max}} = 10^{-2}, 10^{-1}, 10^1, 10^0$ and $q=2.5, 3, 3.5$. As before, the black dashed line shows the critical metallicity of filament formation, namely where $\rho_{\rm{p,max}}/\rho_{\rm{g}} = 100$.

Given a maximum Stokes number of $\tau_{\rm{s,max}} = 10^{-2}$ (first panel), both the system with $q=3$ and $q=3.5$ evolve at about the same rate for about the first $170$ orbits. Then, the $q=3$ system overtakes the density growth and the maximum particle density crosses the dashed line at approximately $175$ orbits. In the case of the system with less mass in larger solids ($q=3.5$), the filament formation begins about $25$ orbits later. 

Given our nominal Stokes number range of $\tau_{\rm{s}} = 10^{-3} - 10^{-1}$ however, the maximum solid density evolution follows the same trend independent of the size exponent, as shown in the second panel of Fig. \ref{compare_surfdens_tau}. The critical metallicity is reached much earlier compared to the case of $\tau_{\rm{s,max}} = 10^{-2}$, after about 100 orbits.

Moving on to the larger particle size range of $\tau_{\rm{s}} = 10^{-3} - 10^{0}$, the metallicity threshold for filament formation is reached after approximately 120 orbits given $q=3$ and about 50 orbits later given $q=3.5$, as shown in the third panel of Fig. \ref{compare_surfdens_tau}. Based on the comparison between all size distribution exponents, and maximum Stokes numbers, it is the run with $q=3$ and $\tau_{\rm{s, max}} = 10^{0}$ that is the most successful in forming permanent filaments early on.

Finally, given the largest size range of $\tau_{\rm{s}} = 10^{-3} - 10^{1}$, filament formation starts almost simultaneously independent of $q$. The two curves on the bottom panel of Fig. \ref{compare_surfdens_tau} show that the maximum solid densities evolve differently once $Z_{\rm{crit}}$ is reached.

We compare the critical metallicities reached in our systems as a function of particle number in Fig. \ref{Z_clump}. We see on the second and third panels that given the smallest maximum Stokes number of $10^{-2}$, filaments form once the metallicity has increased significantly. When comparing $q=3$ and $q=3.5$ for the same particle range of $\tau_{\rm{s}} = 10^{-3}-10^{-2}$, we see that the latter case corresponds to higher $Z_{\rm{crit}}$ values. This is not surprising, as the system with an exponent of $q=3$ contains more mass in larger species, which are more successful in driving the streaming instability. As found by \cite{Bai2010}, it is particles of $\tau_{\rm{s}} \approx 10^{-2}-10^{-1}$ that actively participate in the streaming instability and thus aid the formation of filaments. In the case of our nominal size range of $\tau_{\rm{s}} = 10^{-3}-10^{-1}$, similar $Z_{\rm{crit}}$ values are reached independent of whether $q=2.5, 3$ or $3.5$. We also find that changing the initial particle scale height to $H_{\rm{p, init}}=1$, such that the particle distribution is unstratified within our simulation domain, results in a similar critical metallicity value as in the case of the stratified system with the same parameters. We vary the initial metallicity as well and see that $Z_{\rm{init}}=0.005$ corresponds to a $Z_{\rm{crit}}$ that agrees well with that of the same system with $Z_{\rm{init}}=0.01$. Once the maximum Stokes number is increased to $10^0$, there is quite a large change in the resulting critical metallicity when comparing the systems with $q=3$ and $q=3.5$, as shown in the second and third panels. However, when $\tau_{\rm{s, max}}$ is increased to $10^1$ the $Z_{\rm{crit}}$ values are nearly identical for both size exponents that we examined. Overall, there is no clear trend for how the combination of the particle size distribution exponent and the Stokes number range affect the outcome of the evolution.

\begin{figure}[!th]
\centering
\includegraphics[width=0.7\columnwidth]{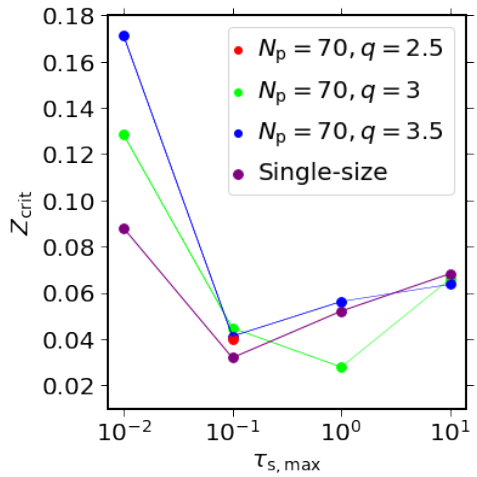}
\caption{Critical metallicity for filament formation, $Z_{\rm{crit}}$, as a function of particle number and maximum Stokes number with $H_{\rm{p,init}}=0.015$ and $Z_{\rm{init}}=0.01$. We show the effect of changing $q$, while keeping the particle number fixed as $N_{\rm{p}} = 70$ per cell. We also show the single-species values from Fig. \ref{Zclump_single_compare} for comparison and see that the systems with more mass in large particles ($q=3$) correspond to lower $Z_{\rm{crit}}$ values than those with $q=3.5$. This is however not the case when $\tau_{\rm{s,max}} = 10^{-1} \mbox{ or } 10^1$. Given these maximum Stokes numbers (or Stokes numbers given the single-species case), the critical metallicity values are similar, independent of the size distribution exponent.}
\label{Z_clump_tau}
\end{figure}

Figure \ref{Z_clump_tau} serves as comparison between the effect of $\tau_{\rm{s,max}}$ and $q$ (given $N_{\rm{p}} = 70$ per cell). We also show here how our single-species results presented in Fig. \ref{Zclump_single_compare}, compare to these values. It is clear that in the case of $\tau_{\rm{s, max}}=10^{-2}$ the critical metallicity threshold approximately doubles as we go from the single-species case to $q=3.5$. This transition can be explained by decreasing the amount of large particles in the system and having less active species available. In the case of $\tau_{\rm{s, max}} = 10^{-1}$, changing the size distribution has little effect on $Z_{\rm{crit}}$. This could be due to the fact that the size dependency for the streaming instability is generally weak in this parameter space. For $\tau_{\rm{s, max}} > 10^{-1}$, changing $q$ does not contribute to significant changes in $Z_{\rm{crit}}$, as in these systems moving mass to smaller species is still keeping a significant fraction of the mass in active particles of $\tau_{\rm{s}} \approx 10^{-2}-10^{-1}$ \citep{Bai2010a}. So overall, there is no clear trend for how changing the Stokes number range changes the efficiency of filament formation.

\section{Conclusion}

In this paper we have studied how the multi-species streaming instability behaves given change in the following parameters: particle number, particle size range and number density distribution exponent. In order to compare how each of these parameters influence the outcome, we compare the critical metallicity where filament formation begins. We point out that these $Z_{\rm{crit}}$ values are not necessarily absolute, given the relatively low resolution of our simulations, but rather serve as a way to compare the effects of the previously mentioned parameters on particle clumping. In Figs. \ref{Z_clump} and \ref{Z_clump_N} we summarize these results and show that varying the particle number does not lead to a significant change in the efficiency of filament formation, given our chosen criteria. We note however, that as Fig. \ref{filaments_q3p5_compare} shows, whether the formed filaments merge or not, can vary as a function of particle number. On the other hand, if self-gravity was included, these filaments would likely form planetesimals around the time the critical metallicity is reached. We also found that there is no correlation between how changing the size distribution exponent affects the outcome.

To understand the dynamics of small particles well, 3D simulations are needed \citep{Bai2010} and in the future the systems discussed in this paper should be tested in 3D as well. In this work, we are limited by fixed resolution box size and somewhat short simulation times. However, keeping such numerical parameters fixed makes comparisons between others more straightforward.

Our results imply that given the realistic case of multiple particle species, the particle size distribution, initial metallicity and initial particle scale height do not have a significant effect on the outcome of the efficiency and timing of filament formation, given our chosen criteria. Our systems reach similar dust-to-gas ratios (see Fig. \ref{compare_epsilon}) - which is set naturally as a consequence of the particle sedimentation - and thus reach the critical midplane density for clumping around the same time. Another result of particle sedimentation is universally high-density conditions in the midplane. These are conditions under which linear multi-species streaming instability models also find short growth timescales \citep{Krapp2019, Paardekooper2020, Paardekooper2021, McNally2021}. As a consequence, the multi-species streaming instability could indeed be successful in forming solid clumps.

\begin{acknowledgements}

We thank the anonymous referee for their comments that helped improve the manuscript. N.S. was funded by the ''Bottlenecks for particle growth in turbulent aerosols'' grant from the Knut and Alice Wallenberg Foundation (2014.0048). N.S. is thankful to Daniel Carrera for useful discussions. A.J. thanks the Swedish Research Council (grant 2018-0486), the Knut and Alice Wallenberg Foundation (grant 2017.0287) and the European Research Council (ERC Consolidator Grant 724687-PLANETESYS) for research support. The simulations were performed on resources provided by the Swedish National Infrastructure for Computing (SNIC) at LUNARC in Lund University.

\end{acknowledgements}

\bibliographystyle{aa}
\bibliography{BigMultiRefs}

\begin{appendix}

\section{Gas removal time}
\label{App1}

Here, we test whether the speed at which the gas is removed from the system matters for when the critical metallicity of filament formation is reached. Other than the nominal gas removal timescale of 50 orbits, we also study systems where the gas density is halved every 100 orbits instead. In Table \ref{gas_rem}, we compare the critical metallicity for both cases. We expect that the resulting $Z_{\rm{crit}}$ values do not significantly depend on the removal time. We see that $\Delta Z_{\rm{crit}} \approx 0.02$ given a 50 orbit delay between $t_{\rm{rem}}=50$ orbits versus $t_{\rm{rem}}=100$ orbits. Overall, as also seen in Fig. \ref{Zclump_single_compare} our nominal removal timescale of $t_{\rm{rem}}=50$ orbits serves as a higher limit for the $Z_{\rm{crit}}$ of filament formation.

\begin{table}[h!]
\centering
\caption{Gas removal timescale}
\label{gas_rem}
\begin{tabular}{>{$}c<{$}>{$}c<{$}>{$}c<{$}}
\hline
\hline
 t_{\rm{rem}} & t_{\rm{crit}} & Z_{\rm{crit}}\\
  (\text{orbits}) & (\text{orbits}) & \\
  \hline
 50 & \sim 113 & \sim 0.048 \\
 100 & \sim 148 & \sim 0.027 \\
\hline
\hline
\end{tabular}
\end{table}

\begin{figure}[!h]
\centering
\includegraphics[width=1\columnwidth]{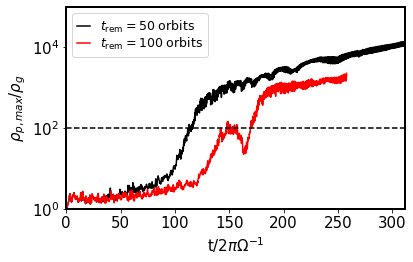}
\caption{Maximum solid density evolution as a function of time comparing the effect of the time of gas removal, such that the gas density is halved every 50 or 100 orbits, respectively. Here, $q=3.5$, $\tau_{\rm{s}} = 10^{-3}-10^{-1}$ and both systems have $N_{\rm{p}} = 7$ particles per cell. The dashed line corresponds to $\rho_{\rm{p,max}}/{\rho}_{\rm{g}} = 100$, which represents the critical maximum solid density needed for filament formation. There is a difference of about 0.02 in the critical metallicity, depending on the gas removal time.}
\label{compare_surfdens_gas}
\end{figure}

\clearpage

\section{Single species solid density}
\label{Appendix1}

\begin{figure*}[th!]
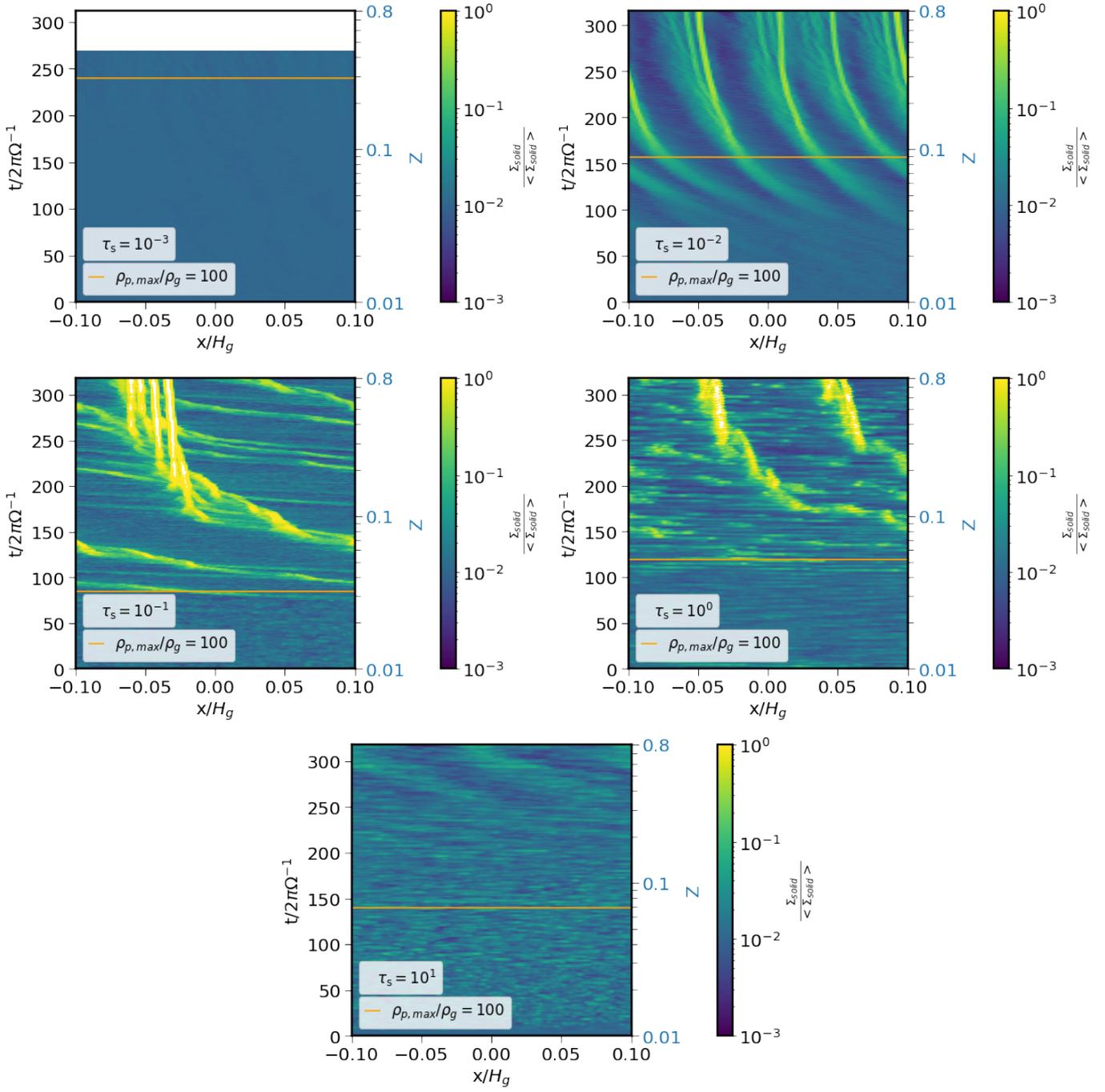

\centering
\includegraphics[width=\columnwidth]{filaments_t0p001_contour.pdf}
\includegraphics[width=\columnwidth]{filaments_t0p01_contour.pdf}
\includegraphics[width=\columnwidth]{filaments_t0p1_contour.pdf}
\includegraphics[width=\columnwidth]{filaments_t1_contour.pdf}
\includegraphics[width=\columnwidth]{filaments_t10_contour.pdf}
\caption{Space-time plot of runs with $\tau_{\rm{s}} = 10^{-3}, 10^{-2}, 10^{-1}, 10^0 \text{ and } 10^1$ with respect to radial coordinate, $x$ and orbital time. The y axis on the right-hand side shows the metallicity, Z. The colors correspond to the solid surface density and the orange line to $\rho_{\rm{p,max}}/{\rho}_{\rm{g}} = 100$. This criteria fits well with the start of filament formation in all systems, except in the first panel, where $\tau_{\rm{s}} = 10^{-3}$. Here, the limit is reached without the formation of any visible filaments.}
\label{single_space_time}
\end{figure*}

In Fig. \ref{single_space_time}, we show the space-time figures of the runs for particle sizes of $\tau_{\rm{s}} = 10^{-3},10^{-2},10^{-1},10^{0},  \text{ and } 10^{1}$. The orange lines mark the critical metallicity where the $\rho_{\rm{p,max}}/\rho_{\rm{g}}$ limit is met. All five runs meet this criteria and for a more detailed discussion see Section \ref{SingleSection}.

\begin{figure}[!h]
\centering
\includegraphics[width=1\columnwidth]{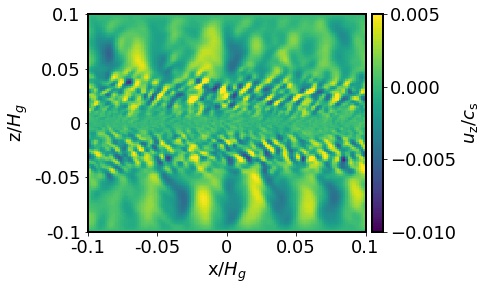}
\caption{Vertical component of the gas velocity as a function of sound speed of the system with $\tau_{\rm{s}} = 10^{-3}$ at $t \sim 260 \mbox{ } 2\pi\Omega^{-1}$. Above the sedimenting particle layer, we find the presence of vertically-elongated structures in $u_{\rm{z}}$.}
\label{uz_0p001}
\end{figure}

In Fig. \ref{uz_0p001}, we show a snapshot of the vertical component of the gas velocity as a function of sound speed at $t \sim 260 \mbox{ } 2\pi\Omega^{-1}$. Here, the particles are in the process of settling and form a thick particle layer around the midplane. We find the presence of vertically-elongated structures in the vertical component of the gas velocity above the diffused particle layer \citep{Lin2021}.
\clearpage

\end{appendix}

\end{document}